\documentclass{aa}

\usepackage{graphicx}
\usepackage{txfonts}
\usepackage{longtable}
\usepackage{amsmath}
\usepackage{float}
\usepackage{multirow}
\usepackage{subfigure}
\usepackage{rotfloat}
\usepackage{xcolor}

\usepackage[]{hyperref}
\usepackage{breakurl}

\usepackage{array}
\usepackage[switch]{lineno}
\usepackage{booktabs}
\usepackage{lscape}

\begin{document}

\title{Second-scale periodicity in FRB 20201124A: coincident with critical RM evolution epochs and its possible implications}

\author{Wen-Long~Zhang\inst{1,2,3,4}
\and Chen-Ran~Hu\inst{5}
\and Chen~Du\inst{5}
\and Zhen-Yin~Zhao\inst{5}
\and Wang-Chen~Xue\inst{2,6}
\and Wen-Jun~Tan\inst{2,6}
\and Shuang-Xi~Yi*\inst{1}
\and Shao-Lin~Xiong*\inst{2}
\and Fa-Yin~Wang\inst{5}
\and Li-Ming~Song\inst{2}
\and Cheng-Kui~Li\inst{2}
\and Shuang-Nan~Zhang*\inst{2,6}
\and Chen-Wei~Wang\inst{2,6}
\and Sheng-Lun~Xie\inst{2,7}
\and Xiao-Fei~Dong\inst{5}
\and Yong-Feng~Huang\inst{5,8}
}

\institute{$^1$ School of Physics and Physical Engineering, Qufu Normal University, Qufu 273165, China\\
\email{yisx2015@qfnu.edu.cn}\\
$^2$ State Key Laboratory of Particle Astrophysics, Institute of High Energy Physics, Chinese Academy of Sciences, Beijing 100049, China\\
\email{xiongsl@ihep.ac.cn; zhangsn@ihep.ac.cn}\\
$^3$ Purple Mountain Observatory, Chinese Academy of Sciences, Nanjing 210023, China.\\
$^4$ School of Astronomy and Space Sciences, University of Science and Technology of China, Hefei 230026, China\\
$^5$ School of Astronomy and Space Science, Nanjing University, Nanjing 210023, China\\
$^6$ University of Chinese Academy of Sciences, Chinese Academy of Sciences, Beijing 100049, China\\
$^7$ Institute of Astrophysics, Central China Normal University, Wuhan 430079, China\\
$^8$ Key Laboratory of Modern Astronomy and Astrophysics
(Nanjing University), Ministry of Education, China}

\markboth{Second-scale periodicity in FRB 20201124A: coincident with critical RM evolution epochs and its possible implications}{}

\abstract{
Second-timescale periodicity ($\sim$1.7\,s) has been reported in the repeating fast radio burst FRB\,20201124A during two epochs, MJD\,59310 and MJD\,59347. We find that these two epochs coincide with the days on which the Faraday rotation measure (RM) of the source passes through the transition between the quasi-steady and the strongly varying phases that are predicted by the magnetar/Be-star binary model. To examine whether this coincidence could arise by chance, we re-fitted the secular RM evolution within the same binary framework and estimated the probability of chance coincidence, $P_{\rm CC}\approx2.5\%$. Independently, and in order to cross-validate the reported periodicity, we carried out a coherent joint period search on the bursts of the two epochs, assuming a simple linear evolution of the spin period; we obtain $P_0=1.70603$\,s and $\dot{P}=6.1\times10^{-10}$\,s\,s$^{-1}$ at the reference epoch, with a Gaussian significance of $\sim3.9\sigma$, consistent with the previous measurement. We propose that the two periodic epochs correspond to passages of the magnetar through the decretion disk of the Be-star companion. The captured disk material suppresses low-latitude magnetospheric emission, so that the polar-cap emission dominates and the spin period becomes detectable; alternatively, because the disk material carries substantial angular momentum, accretion increases the magnetic obliquity and thereby amplifies the periodic modulation of the polar-cap emission. The synchronization between the burst periodicity and the magneto-ionic transition points to a link between burst activity and the dynamical evolution of the source environment. The details of the binary-model fit and of the period search are given in Appendices~\ref{app:binary} and~\ref{app:period}.
}

\authorrunning{Zhang et al.}
\titlerunning{Second-scale periodicity in FRB 20201124A}
\maketitle
\nolinenumbers

\section{Introduction}

Fast radio bursts (FRBs) are millisecond-duration highly energetic radio transients of
extragalactic origin \citep{Xiao2021,2023RvMP...95c5005Z,2025arXiv251019143K}. Although discovered more than a decade ago
\citep{2007Sci...318..777L}, their emission mechanism remains an open question. The detection of 
FRB 20200428A from a Galactic magnetar SGR J1935+2154 \citep{2020Natur.587...54C,2020Natur.587...59B} provided compelling evidence that at least a subset of FRBs can originate from magnetar activity. FRB 
20200428A was observed during an active bursting episode of SGR J1935+2154, which exhibited an 
intense X-ray burst storm. The association between FRB 20200428A and a particular X-ray burst
\citep{2021NatAs...5..378L,2020ApJ...898L..29M, 2021NatAs...5..372R,2021NatAs...5..401T} suggests that these X-ray events may constitute a distinct subclass
with unique physical properties \citep{2023RAA....23k5013Z,2024ApJ...967..108X}. Follow-up 
observations of the sole confirmed Galactic FRB source to date have led to the detection of numerous FRBs from it
\citep{2023SciA....9F6198Z,  2024ApJS..275...39W}. Subsequent X-ray monitoring revealed that the
magnetar undergoes spin acceleration both before and after FRB activity
\citep{2023NatAs...7..339Y,2024Natur.626..500H}.

Generally, FRBs are categorized into repeating and non-repeating (one-off) groups \citep{2021ApJ...911L...3P}. 
Repeaters emit multiple bursts over time, whereas non-repeating sources have only been detected to emit a single burst throughout observational campaigns. The observed properties of repeating FRBs typically require more complex physical mechanisms and environments for interpretation. 
Rotation measure (RM) serves as a key observational quantity that depends on the electron density and the magnetic field along the line-of-sight ($\mathrm{RM} \propto \int n_e B_\parallel dl$). It provides a useful probe of the magnetic environment surrounding FRB sources. A series of related observational studies have been conducted \citep{2018Natur.553..182M,2021ApJ...908L..10H,Anna-Thomas2023}.
The growing number of repeating FRBs exhibiting RM variations along with the emergence of long-period features suggestive of orbital motion has complicated the interpretation of these phenomena \citep{2020Natur.582..351C,2020MNRAS.495.3551R,2020MNRAS.496.3390B, 2022NatCo..13.4382W,2025arXiv250506006X,2025arXiv250510463L}. This complexity has motivated the widespread adoption of binary system models, in which a neutron star serves as the emitter orbiting a companion star \citep{2025ApJ...994L..20Z,2025arXiv251207140W,2025arXiv251212995S}.

Recent applications of these models have effectively explained observations such as the RM sign reversal in FRB 20201124A \citep{2022NatCo..13.4382W} and the  RM evolution and ``RM flare'' in FRB 20220529A \citep{2025arXiv250304727L,2026arXiv260102734D}, offering a compelling explanation for phenomena not readily addressed by isolated magnetar models. 
FRB 20201124A is a notable repeater that has been 
extensively studied due to its active bursting behavior and complex local environment 
\citep{2022ApJ...927...59L,2022Natur.609..685X}.

Recently, \cite{2025arXiv250312013D} reported a second-timescale periodicity in FRB 20201124A detected on two separate days. The findings are confirmed by \cite{2025arXiv251223392S}, 
who also identified similar periodicity on the two days. Specifically, two distinct periodicities in bursts from FRB 20201124A observed by FAST are identified on MJD 59310 and MJD 
59347, respectively. This discovery provides important observational evidence for FRB models. To explain why the second-scale periodic signature, potentially linked to the magnetar's rotation, was only detected on these two days, \cite{2025arXiv250312013D} invoke the quenching of other radiation fields, leaving behind a stable zone whose rotation revealed the periodicity. The low burst rate observed during these days makes this interpretation plausible. Nevertheless, a more compelling explanation still needs to be explored.

We noticed that the periodicity was reported on MJD\,59310 and MJD\,59347, and that these two days coincide with the epochs at which the RM of FRB\,20201124A passes through the transition between its quasi-steady and strongly varying phases. These are precisely the epochs at which the magnetar is expected to cross the decretion disk of its Be-star companion in the binary scenario proposed by \citet{2022NatCo..13.4382W}. This raises the possibility that the appearance of the second-scale periodicity is not accidental, but is instead caused by the interaction of the magnetar with the disk material. Such a coincidence could, however, also arise by chance. In order to quantify it, we re-fitted the secular RM evolution of FRB\,20201124A within the same magnetar/Be-star binary model and estimated the corresponding probability of chance coincidence (Appendix~\ref{app:binary}). Independently, and in order to cross-validate the reported periodicity with an analysis of our own, we searched for a coherent periodic signal in the bursts of the two epochs, assuming a simple linear evolution of the spin period (Appendix~\ref{app:period}). We stress that both analyses are used here as supporting evidence. The main purpose of this paper is to propose and examine the physical picture in which the magnetar crosses the Be-star disk.

The paper is organized as follows. In Section~\ref{sec:methods} we describe the data and the two analyses that support this work, namely the magnetar/Be-star binary fit and the joint coherent period search. In Section~\ref{sec:results} we present the observational evidence that links the two epochs of periodicity to the RM evolution, we quantify the probability of chance coincidence, and we discuss the physical picture of a magnetar crossing the Be-star disk. Our conclusions are given in Section~\ref{sec:conclusions}. The technical details of the two analyses are deferred to Appendices~\ref{app:binary} and~\ref{app:period}, respectively.

\begin{figure}[http]
\centering
\includegraphics[width=\columnwidth]{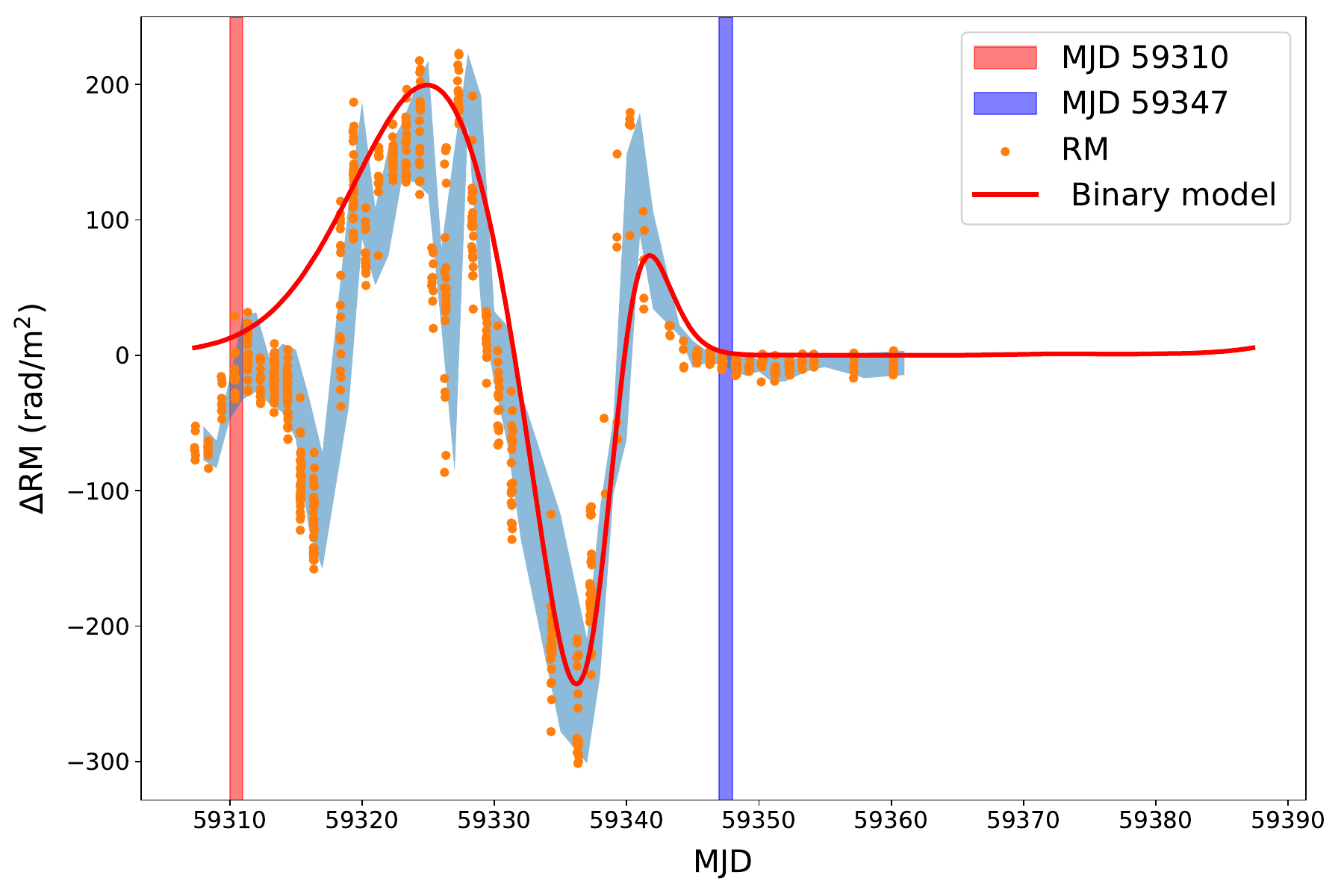}\\
\caption{Evolution of the RM of FRB 20201124A relative to a constant background (from \citep{2022NatCo..13.4382W}). The orange points are the measured RM values and the red curve is the best-fit magnetar/Be-star binary model (Appendix~\ref{app:binary}). The red and blue vertical bands mark MJD\,59310 and MJD\,59347, the two epochs at which the second-scale periodicity is detected.}
\label{RMe}
\end{figure}

\section{Methods}
\label{sec:methods}

We used the FAST observations of FRB\,20201124A. The RM values and the burst arrival times are taken from \citet{2022Natur.609..685X}. The source was monitored over 54 days, and the two epochs at which the second-scale periodicity was reported are MJD\,59310 and MJD\,59347 \citep{2025arXiv250312013D,2025arXiv251223392S}.

In order to describe the secular evolution of the RM, we adopted the magnetar/Be-star binary model of \citet{2022NatCo..13.4382W}, in which a neutron star orbits a Be star and crosses its equatorial decretion disk twice per orbit. As the neutron star moves through the disk, the electron density and the magnetic field strength along the line of sight change, producing the observed RM variations. We fitted the binned RM series with this model using a Markov Chain Monte Carlo (MCMC) method, leaving eight parameters free: the orbital period, the orbital eccentricity, the surface magnetic field strength, the orbital inclination, two viewing-angle components, the background RM, and an additional noise term. The model, the fitting procedure, and the resulting posterior distributions are described in Appendix~\ref{app:binary}.

The epochs of disk crossing are identified with the transition phases of the RM evolution, that is, the intervals in which the first derivative of the fitted RM curve changes sign and the curve approaches the background level. In order to quantify whether the coincidence between these epochs and the two days of periodicity could arise by chance, we defined, following the fitted model, Epoch~1 to cover MJD\,59307--59312 and Epoch~2 to cover MJD\,59343--59348. Each epoch spans from the sign change of the first derivative to the intersection of the best-fit curve with the upper $3\sigma$ background level. We then computed the probability that two randomly selected days out of the 54 days of observation fall within these windows, $P_{\rm CC}$.

Finally, in order to test the reported periodicity with an analysis of our own, we performed a coherent joint period search on the bursts of the two epochs. We assumed a simple linear evolution of the spin period, $P(t)=P_0+\dot{P}(t-t_0)$, folded the burst arrival times in phase with 20 bins, and searched a two-dimensional grid in $P_0$ and $\dot{P}$. The significance of the resulting phase-folded signal was assessed with $10^5$ Monte Carlo simulations of Poisson-distributed arrival times matching the observed number of bursts and time span. The details are given in Appendix~\ref{app:period}.

\section{Results and discussion}
\label{sec:results}

We now examine the relation between the two epochs of periodicity and the secular evolution of the RM. Figure~\ref{RMe} shows the RM of FRB\,20201124A relative to a constant background, together with the best-fit magnetar/Be-star binary model of \citet{2022NatCo..13.4382W} that we have re-fitted to the data (Appendix~\ref{app:binary}). The measured RM varies by several hundred rad\,m$^{-2}$ on a timescale of weeks, with a pronounced excursion between MJD\,59330 and MJD\,59345.

The binary model provides a good description of the basic trend of the RM evolution, with a reduced chi-squared statistic of $\chi^2_{\nu}=0.99$ (Appendix~\ref{app:binary}). The derived orbital period, $P_{\rm orb}=79.51_{-0.91}^{+0.81}$ days, is consistent with the $\sim$80-day period reported by \citet{2022NatCo..13.4382W}. The temporal evolution of the first derivative of the fitted curve is presented in Appendix~\ref{app:binary}. The periodic features appear on MJD\,59310 and MJD\,59347. These two days occur near the transition phase in the first-derivative evolution, at the transition between the quasi-steady and the strongly varying phases of the RM.

Following the procedure described in Section~\ref{sec:methods}, we estimate the probability of chance coincidence (Pcc) for the alignment between these two periodic epochs and the RM transition intervals. The probability is given by
\begin{equation}
P_{\rm CC} \approx \frac{6 \times 6}{C_{54}^{2}} = \frac{36}{1431} \approx 2.5\%.  
\end{equation}
Although this chance coincidence is not very low, owing to the limited data volume and the large uncertainty of the RM transition epochs, the physical mechanism proposed below strengthens the reliability of this temporal alignment. We further note that the two days coincide with points that are almost symmetrically distributed around the midpoint of a sustained declining phase in the local RM evolution that intersects the baseline level (BKG), as illustrated by the orange shaded regions in Appendix~\ref{app:binary}. Their separation is 37 days, close to half of the best-fit orbital period, $P_{\rm orb}/2\simeq39.8$ days. This is what is expected if the two epochs correspond to the two passages of the magnetar through the Be-star disk within a single orbit, one before and one after periastron. Such a symmetry provides additional supporting evidence that the correspondence may in fact reflect an underlying physical regularity rather than mere chance.

The temporal coincidence discussed above is physically meaningful only if the periodicity itself is real. Both the original detection \citep{2025arXiv250312013D} and the independent re-analysis of \citet{2025arXiv251223392S} indicate that it is. In order to test the signal with an independent method, we performed a coherent joint period search on the FAST bursts of the two epochs, assuming a simple linear evolution of the spin period (Appendix~\ref{app:period}). We obtain $P_0=1.70603$\,s at the reference epoch $t_0=\mathrm{MJD}\,59310.34008104$ and $\dot{P}=6.1\times10^{-10}$\,s\,s$^{-1}$, with a Gaussian significance of $\sim3.9\sigma$. Both the period and its derivative are consistent with the values reported by \citet{2025arXiv250312013D} and lie in the range expected for a magnetar. The bursts of the two epochs cluster in phase, whereas the bursts detected during the rest of the campaign are distributed almost uniformly. This indicates that the emission is phase-locked and likely originates from the polar cap.

The temporal evolution of RM provides critical 
diagnostics for the magnetoionic environment during transition states. The time derivative 
of RM can be expressed as
\begin{equation}
\label{eq3}
\dot{\rm RM} \propto \int_{\rm los} \left( \dot{n}_e B_{\parallel} + n_e \dot{B}_{\parallel} \right) \frac{{\rm{d}}l}{{\left[1+z(l)\right]}^2},
\end{equation}
where $\dot{\rm RM}$ reflects combined contributions from electron density variations 
($\dot{n}_e$) and evolving magnetic field component along the line-of-sight 
($\dot{B}_{\parallel}$). Significant RM variability more likely originates from the source 
environment, such as magnetic reconnection or transient plasma injection. Accordingly, the 
observed synchronization between periodic emission episodes and RM state transitions may 
originate from magneto-plasma evolution proximal to the emission region.

We point out that the observed temporal coincidence between the periodicity epochs and the RM transition days can be explained by the evolution of the emission geometry. During the early and late epochs of a burst-active episode, in particular on MJD\,59310 and MJD\,59347, the radiation could be predominantly confined to the polar-cap region. In contrast, the intermediate epochs exhibit additional emission from non-polar magnetic latitudes (e.g. close field-line regions), which dilutes the periodic modulation and obscures the detectability of the spin period.

\citet{2022Natur.609..685X} first reported the secular RM evolution of FRB 20201124A, later interpreted by \citet{2022NatCo..13.4382W} as signatures of a 
magnetar/Be-star binary system. This framework
\citep{2022NatCo..13.4382W,Zhao2023} exhibits emission variability similar to
supergiant fast X-ray transients (SFXTs) \citep{2006ESASP.604..165N}, although distinct in the spectral regime. \cite{2022RAA....22f5012Z} suggests that the dynamics of the system may also be governed by magnetohydrodynamic interactions between the compact object and the donor star through the Roche lobe.
In our binary framework, the material in the disk surrounding the Be star naturally possesses substantial angular momentum. This readily supplies the kind of accreting material with angular momentum that is required to increase the misalignment between the magnetic and rotational axes, as discussed above, thereby offering a natural and self-consistent explanation of the phenomenon.

\begin{figure}[http!]
\centering
\includegraphics[width=\columnwidth]{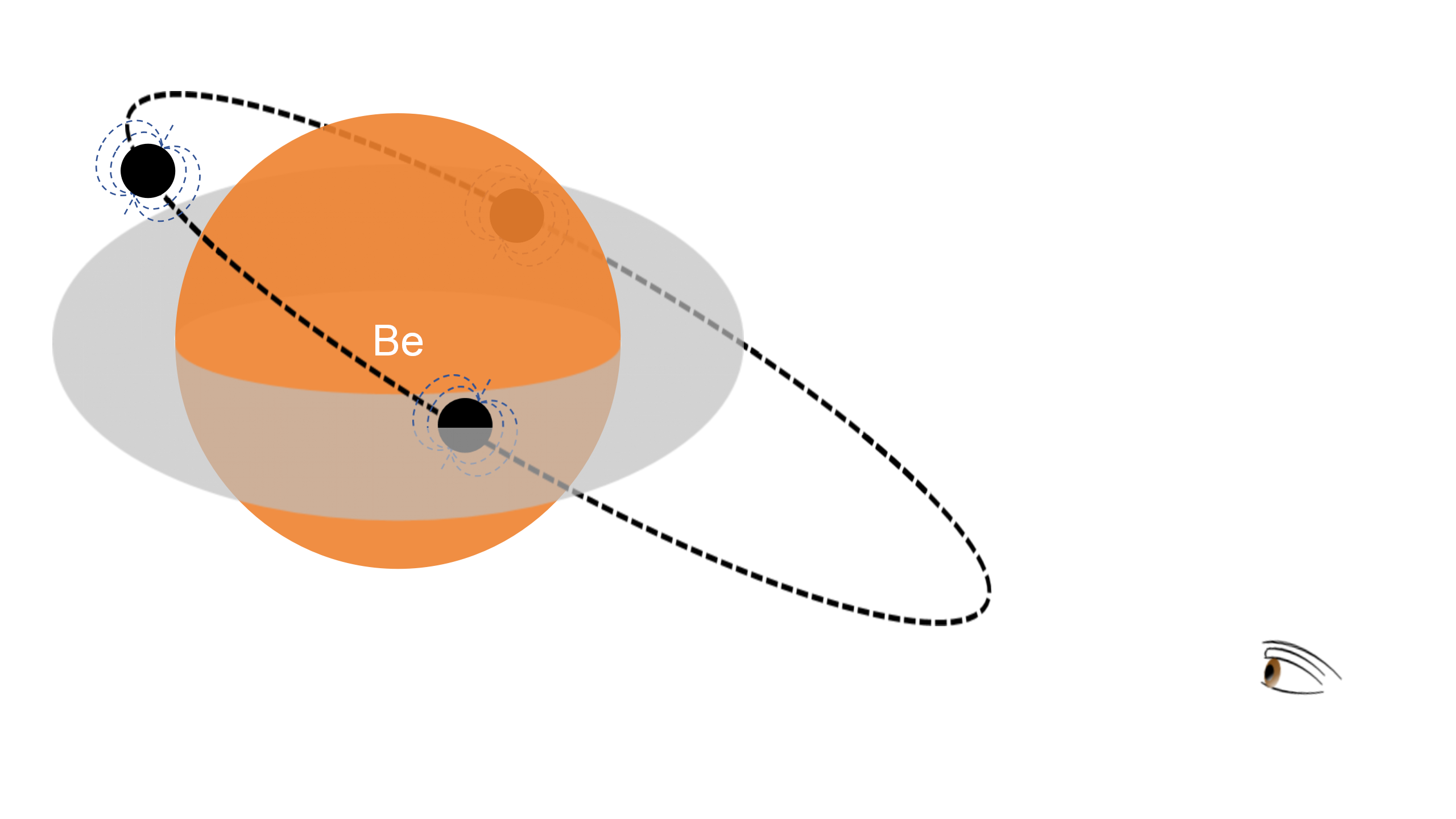}\\
\caption{A schematic diagram depicting the magnetar/Be-star binary system. The two orbital positions of the magnetar with respect to the Be-star disk demonstrate its relative geometric alignment corresponding to the two special epochs.}
\label{Mag_Be}
\end{figure}

Recent work by \citet{2025arXiv250318651Y} proposed the magnetar to be dominant for FRB generation in a binary system. Orbital motion modulates the integrated product of line of sight of the electron
density and magnetic field component, producing the RM evolution signature shown in
Fig.~\ref{RMe} (red curve). We suggest that the observed enhancement and reduction of the RM in the
transition state on MJD 59310 and 59347 correspond to the magnetar passing through the wind disk of the companion star, as depicted in
Fig.~\ref{Mag_Be}. In this picture, a similar periodic signature would be expected to appear whenever the source is observed in a comparable RM pattern during subsequent observations.

The observed multi-timescale behavior in Aquila X-1 
\citep{1991ApJ...375..696S,1998ApJ...499L..65C} provides a compelling analogy for 
understanding periodic phenomena in compact binaries. This prototypical X-ray binary
exhibits two characteristic timescales. The first is a 1.3-day period from orbital modulation of the binary system \citep{1976MNRAS.176P..19W}. The second is a kHz quasiperiodic
oscillation that likely reflects the neutron star spin frequency and is strongly correlated with the accretion process, known as the 
``propeller'' effect \citep{1998ApJ...494L..71Z,1998ApJ...495L...9Z}. Similar 
accretion-regulated periodicity could emerge in the FRB 20201124A source. The markedly low burst rate recorded in these two days offers additional evidence to support the proposed mechanism. As demonstrated by 
\citet{2025arXiv250318651Y}, magnetar encounters with the Be star's decretion disk trigger 
propeller-phase activation through accretion. The temporal synchronization between 
disk-crossing events and burst periodic active epochs can support the source's similarity to 
Aquila X-1\textquoteright s accretion-rotation coupling, albeit operating in distinct energy 
regimes (radio vs. X-ray). After crossing the disk, the combined effects of the 
multi-polar magnetic field recovery and residual plasma interactions naturally generate 
multiple new acceleration fields on the magnetar's surface, resulting in a transient 
enhancement of the burst rate, which can be seen as shown in Fig.1 of 
\cite{2025arXiv250312013D}. For neutron stars, the manifestation of 
multipole field effects depends on whether the location of FRB emission is 
within the light cylinder. \citet{2025Natur.637...48N} conclude that the 
radiation region is located within or just beyond the magnetosphere based on 
local plasma scintillation, which makes the link between the FRB sources and 
multipole fields more plausible, especially for the source discussed in this 
work with $P_0 = 1.70603\, \text{s}$ and $\dot{P} = 6.1 \times 10^{-10} 
\, \text{s}\, \text{s}^{-1}$. We suggest that multipole field effects can be 
considered in the mechanisms of FRB generation, unlike other mechanisms  \citep{2019MNRAS.485.4091M,2020ApJ...889..135L}.

Recent observations of FRB 20201124A provide additional context for interpreting our results. 
\citet{Bilous2025} reported a frequency-dependent activity window, which naturally aligns with our model of a partially obscuring circumstellar disk from a Be-star companion; the disk's opacity is frequency-dependent, attenuating lower frequencies more strongly. 
The claimed persistent radio source (PRS) association by \citet{Bruni2024} warrants consideration within our binary framework. 
While the origin of the PRS remains uncertain, it could arise from a magnetar wind nebula \citep{Margalit2018,Zhao2021} or persistent emission due to ongoing magnetar-disk interactions, particularly during periastron passages or disk-crossing events. 
Notably, our model predicts enhanced burst activity following disk crossings (see Fig. 1 of \citet{2025arXiv250312013D}), which may correlate with PRS variability. 
Finally, \citet{Main2023} identified a Galactic scintillation screen for this source. 
Although they did not explore orbital-phase dependence, our model predicts that scintillation properties such as scintillation velocity and timescale should exhibit modulation over the orbital period. 
Such orbital-phase-dependent scintillation, if detected in future campaigns, would provide compelling evidence for the binary geometry proposed here.

\section{Conclusions} 
\label{sec:conclusions}

In this work we investigate the possible connection between the $\sim$1.7\,s periodic signal reported by \cite{2025arXiv250312013D} in FRB\,20201124A on MJD\,59310 and MJD\,59347 and the transitional phase of its RM evolution, with the aim of identifying its physical origin. We first noticed that these two days coincide with the epochs at which the RM evolution, as described by the magnetar/Be-star binary model of \citet{2022NatCo..13.4382W}, passes through the transition between the quasi-steady and the strongly varying phases. In order to quantify whether this alignment could arise by chance, we re-fitted the secular RM evolution within the same binary framework. We obtained an orbital period $P_{\rm orb}=79.51_{-0.91}^{+0.81}$ days and a reduced chi-squared statistic $\chi^2_{\nu}=0.99$, and we estimated a probability of chance coincidence $P_{\rm CC}\approx2.5\%$ for the alignment of the two periodic epochs with the RM transition intervals (Appendix~\ref{app:binary}).

Independently, and in order to cross-validate the reported periodicity, we performed a joint coherent period search on the bursts of the two epochs under the assumption of a simple linear evolution of the period. We obtained $P_0 = 1.70603$\,s, $\dot{P} = 6.1\times10^{-10}$\,s\,s$^{-1}$, and a Gaussian significance of $\sim3.9\sigma$, consistent with the results of \citet{2025arXiv250312013D} (Appendix~\ref{app:period}).

The two analyses summarized above are used here as supporting evidence for the main proposal of this work. We propose that the two epochs of second-scale periodicity correspond to passages of the magnetar through the decretion disk of its Be-star companion. We then discussed the emergence of this second-timescale periodicity within the framework of a single magnetar as the source of FRB 20201124A. In this scenario, the periodic signal could arise if accretion of surrounding material quenches multipolar magnetic field radiation at low latitudes, temporarily allowing radiation from the polar cap region to dominate and thus revealing the spin period. Alternatively, accretion could increase the angle between the magnetic and rotation axes. This would enhance the visibility of the spin modulation if the accreted material carries sufficient angular momentum.

The disk material surrounding the Be star naturally provides accreting material with the required angular momentum, which is why we introduced the magnetar/Be-star binary model. Using the model and code presented by \cite{2022NatCo..13.4382W}, we relaxed free parameters to obtain a satisfactory fit to the overall RM evolution trend. Although not a smoking-gun signature, the two days in question show consistency with the transitional phase in the RM fit, and this possibility cannot be ruled out.

In the binary framework, the two epochs likely correspond to points where the magnetar's orbit intersects the Be-star's disk. During disk crossing, plasma flows onto the magnetar and could suppress multipolar field radiation over much of the surface, significantly reducing the FRB burst rates clearly seen in Fig.1 of \cite{2025arXiv250312013D}. This would allow polar-cap emission to dominate, making the spin period detectable. Alternatively, because the disk plasma inherently possesses angular momentum, accretion could increase the magnetic obliquity, in contrast to Roche-lobe accretion which tends to reduce it. Both pathways naturally explain the appearance of the spin period.

From the perspective of binary orbital periodicity, similar activity is expected to recur at the same orbital phase. However, subsequent observations reveal that the burst behavior and the evolution of parameters such as the rotation measure during later active episodes differ significantly from those observed in this phase. This discrepancy may be attributed to precession of the binary orbit or the accretion disk, leading to substantial variations in the observational characteristics of the bursts, as also discussed in related works \citep{2025arXiv251212995S,2026arXiv260102734D}. To date, no such recurrence has been observed. Nevertheless, we anticipate that similar features may be detected in other analogous sources in the future. Further extensive monitoring on FRB 20201124A is also solicited.

While both the isolated magnetar and binary models can qualitatively account for the observations, the binary model offers a more self-consistent explanation. Although current evidence does not strongly exclude other physical models, and the coincidental alignment remains only suggestive, this work provides a novel theoretical perspective on the origin of transient periodic signatures in repeating FRBs.

\begin{acknowledgements}
We thank Di Xiao, Ye Li, Ming-Yu Ge, Shu-Xu Yi, Xiao-Bo Li and Shu Zhang for their helpful discussions. This work is supported by the National Natural Science Foundation of China (Grant No. 12673051, 12494572, 12494575, 12273042, 12333007, 12233002), the Strategic Priority Research Program, the Chinese Academy of Sciences (Grant No. XDA30050000, XDB0550300), the National Key R\&D Program of China (2021YFA0718500), Shandong Provincial Natural Science Foundation (ZR2025MS47), the Natural Science Foundation of Jiangxi Province of China (20242BAB26012) and Manned Spaced Project (CMS-CSST-2021-A12). XFD acknowledges the Postgraduate Research \& Practice Innovation Program of Jiangsu Province (No. KYCX25\_0197). YFH also acknowledges the support from the Xinjiang Tianchi Program.
\end{acknowledgements}

\bibliographystyle{aasjournal}
\bibliography{bibtex.bib}

\begin{appendix}

\section{The magnetar/Be-star binary model}
\label{app:binary}

\subsection{Data and method}

For FRB 20201124A, we adopt the simpler model from \citet{2022NatCo..13.4382W}. A key distinction of our approach is that, while the system is assumed to follow an elliptical orbit, the model utilizes the mean orbital phase derived from a circular orbit to describe the RM evolution, which provides a better fit to the current data. We fit the binned RM data with this model, whose physical setup consists of a Be star ($8M_\odot$, $5R_\odot$) surrounded by a circumstellar disk and a neutron star ($1.6M_\odot$) on an eccentric orbit. As the neutron star traverses different regions of the Be star's disk, it encounters varying electron densities and magnetic field strengths, thereby producing the observed RM variations. It should be noted that this model relies on simplified assumptions. Although certain parameters are left free during fitting, the approach is intended primarily as a functional description of the evolutionary trend.

The model incorporates three key physical components:

\begin{itemize}
    \item Disk density structure: $\rho(r,z) = \rho_0 (r/R_*)^{-n} \exp[-(z/H(r))^2]$, with scale height $H(r) = c_s \sqrt{r^3/GM_*} \cdot r$ for an isothermal disk at $T = 20,000$ K, where $c_s$ is the sound speed;
    \item Dipole magnetic field: $\mathbf{B} = B_0 (\mathbf{r}/R_*) / r^3$ centered on the Be star;
    \item Eccentric orbital motion: The neutron star follows $r = p/(1 + e\cos\theta)$ with $p = a(1-e^2)$, where the semi-major axis $a$ is determined from Kepler's third law.
\end{itemize}

In our Markov Chain Monte Carlo (MCMC) fitting procedure, we fixed several parameters based on typical Be star properties: stellar parameters ($M_* = 8M_\odot$, $R_* = 5R_\odot$), disk density index ($n = 4.0$), base density ($\rho_0 = 3\times10^{-14}$ g/cm$^3$), and disk temperature ($T = 20,000$ K). Notably, the orbital eccentricity $e$ is now included as a free parameter in our analysis, reflecting its important role in shaping the RM variation profile. The eight free parameters in our analysis include:

\begin{itemize}
    \item Orbital period $P_{\rm orb}$;
    \item Orbital eccentricity $e$;
    \item Surface magnetic field strength $B_0$;
    \item Orbital inclination $\phi$;
    \item Viewing angle components $\theta_1$, $\theta_2$;
    \item Background RM term BKG;
    \item Additional noise parameter $E_{\rm quad}$.
\end{itemize}

We employed the MCMC method to generate posterior distributions. The inclusion of eccentricity as a free parameter allows for better characterization of the asymmetric RM variations observed in the data. The likelihood function follows \cite{2025arXiv250506006X}:

\begin{equation}
\mathcal{L} = \prod_{i}\frac{1}{\sqrt{2\pi(\sigma_{i}^{2} + E_{\rm quad}^{2})}} \exp\left[-\frac{1}{2} \frac{({\rm RM}_{i} - {\rm sig}(t_{i}))^{2}}{\sigma_{i}^{2} + E_{\rm quad}^{2}}\right],
\end{equation}
where:
\begin{itemize}
    \item ${\rm RM}_{i}$, $\sigma_{i}$, and $t_{i}$ are the rotation measure, its measurement uncertainty, and the Solar-system barycentric arrival time of the $i$-th burst, respectively;
    \item ${\rm sig}(t_i)$ represents the RM signal predicted by the binary orbital model;
    \item $E_{\rm quad}$ accounts for short-term RM fluctuations exceeding the measurement errors, analogous to the white noise treatment in pulsar timing \citep{2014MNRAS.437.3004L}.
\end{itemize}

To assess the goodness-of-fit, we compute the reduced chi-squared statistic:

\begin{equation}
\chi^{2}_{\nu} = \frac{1}{\nu} \sum_{i} \frac{({\rm RM}_{i} - {\rm sig}(t_{i}))^{2}}{\sigma_{i}^{2} + E_{\rm quad}^{2}},
\end{equation}
where $\nu = N - k$ is the number of degrees of freedom, with $N$ denoting the number of data points and $k$ the number of fitted parameters.

\begin{figure}[http!]
\centering
\includegraphics[width=\columnwidth, angle=0]{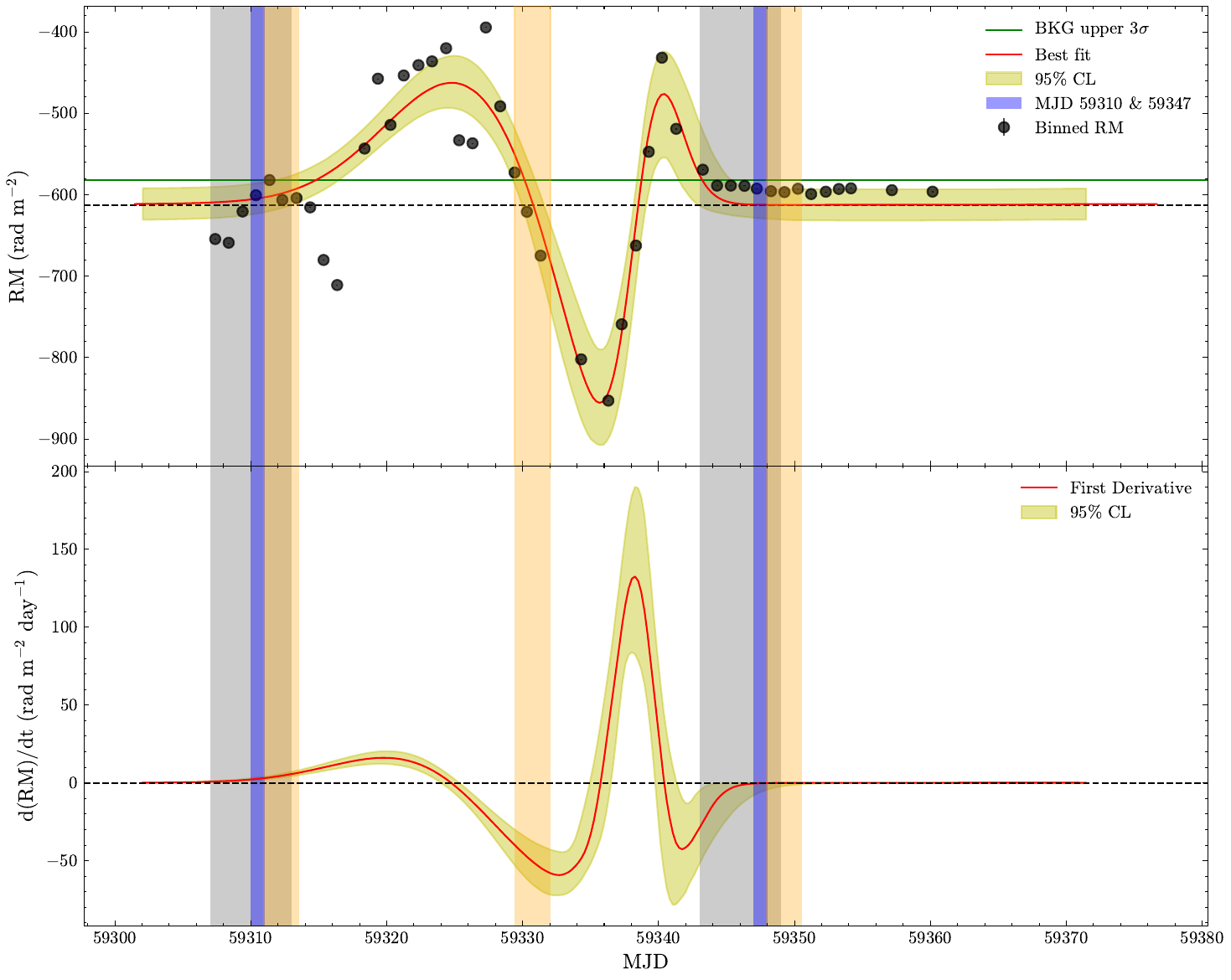}\\
\caption{Upper subplot: RM evolution of FRB 20201124A fitted with the magnetar/Be-star binary model; the middle orange shaded region represents the 3$\sigma$ confidence range of the intersection between the theoretically predicted RM and the background RM in this region, and the other two orange segments indicate the intersection region shifted forward or backward by $\sim$ 18.5 days, which is half the separation between the two days, showing that the periastron position lies almost midway between them. Lower subplot: first derivative of the fitted curve.}
\label{RM_fit}
\end{figure}

\begin{figure*}[http!]
\sidecaption
\includegraphics[width=1.5\columnwidth, angle=0]{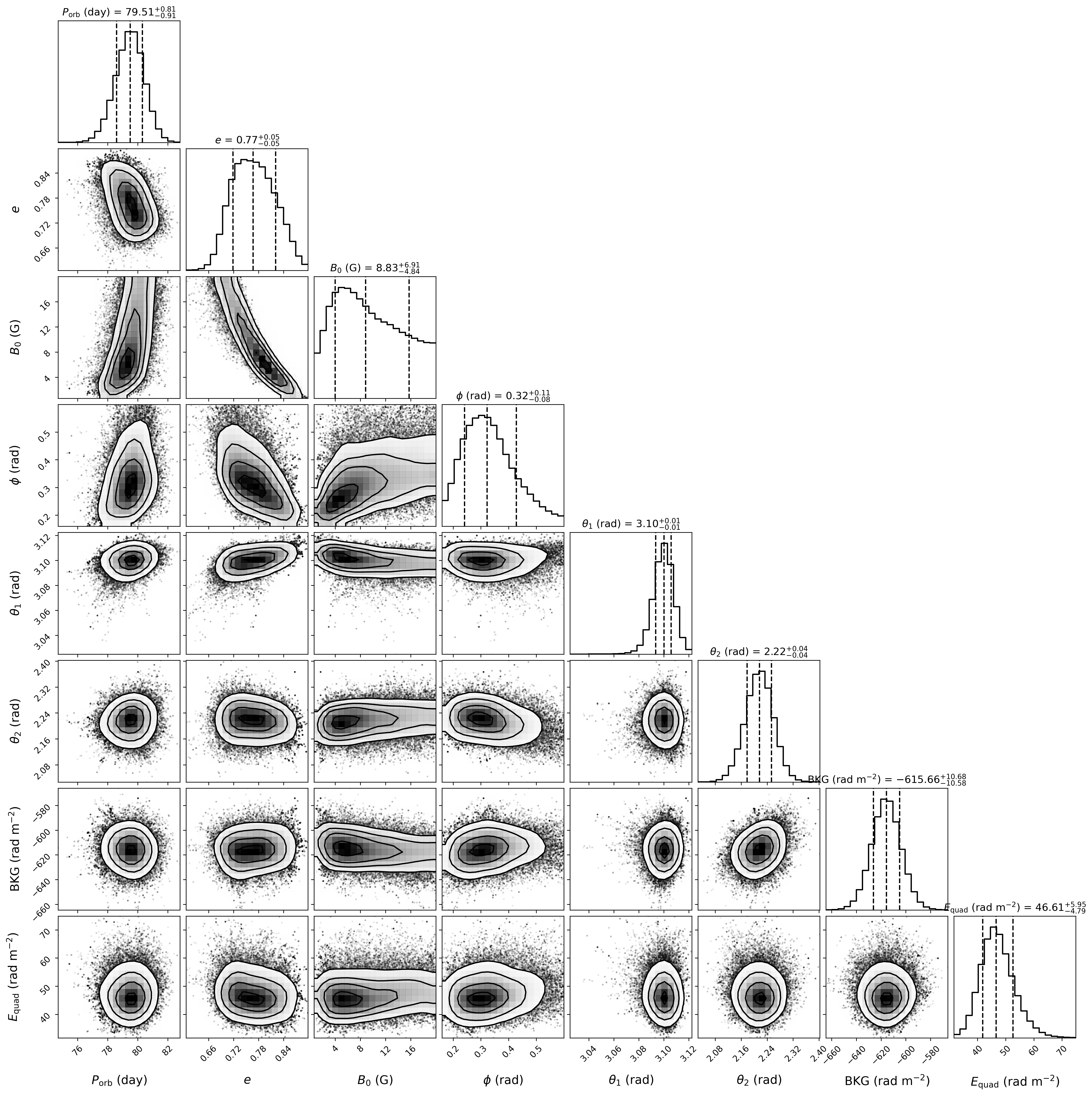}
\caption{Posterior distributions and fitting results of the model parameters, shown at the maximum width allowed by the two-column layout so that the axis labels remain legible. From left to right along the diagonal: $P_{\rm orb}$, $e$, $B_0$, $\phi$, $\theta_1$, $\theta_2$, BKG, and $E_{\rm quad}$.}
\label{corner}
\end{figure*}

\subsection{Results}

The best-fit binary model is shown in Fig.~\ref{RM_fit}, and the posterior distributions of the eight free parameters are shown in Fig.~\ref{corner}. The model reproduces the basic trend of the RM evolution, including the large excursion between MJD\,59330 and MJD\,59345, with a reduced chi-squared statistic of $\chi^2_{\nu} = 0.99$. The derived orbital period, $P_{\rm orb} = 79.51_{-0.91}^{+0.81}$ days, is consistent with the $\sim$80-day period reported by \citet{2022NatCo..13.4382W}. As discussed in Section~\ref{sec:results}, the second-scale periodicity is detected on MJD\,59310 and MJD\,59347. These two days lie close to the epochs at which the first derivative of the fitted curve changes sign.

We caution that the binary model contains a large number of free parameters and should be regarded as a phenomenological description of the RM evolution rather than a unique solution. In particular, the posterior distributions are broad for several parameters, and the model is intended to capture the overall evolutionary trend rather than to provide precise physical constraints. The temporal coincidence between the two periodic epochs and the RM transition phases should therefore be judged primarily from the physical picture developed in Section~\ref{sec:results}, and not from the quality of the fit alone.

\section{The isolated magnetar model and the joint period search}
\label{app:period}

\subsection{Data and method}

In order to cross-validate the second-scale periodicity reported by \citet{2025arXiv250312013D} with an independent analysis, we used the FAST burst arrival times of FRB\,20201124A reported by \citet{2022Natur.609..685X}. Following the isolated-magnetar picture, we modelled the spin period as a linear function of time,
\begin{equation}
\label{eq1}
P(t) = P_0 + \dot{P}\,(t-t_0),
\end{equation}
where $P_0$ is the initial spin period at the reference epoch $t_0$ and $\dot{P}$ its derivative. The rotational phase was obtained by integrating the instantaneous angular frequency $\omega(t)=2\pi\nu(t)$, with $\nu=1/P$:
\begin{equation}
\label{eq2}
\phi(t)=\phi_0+\nu_0\,(t-t_0)+\tfrac{1}{2}\,\dot{\nu}_0\,(t-t_0)^2
+\mathcal{O}\!\left[(t-t_0)^3\right].
\end{equation}
Equation~(\ref{eq2}) is the standard Taylor expansion of the phase in the spin frequency. It remains well defined in the limit $\dot{P}\to0$, where it reduces to $\phi(t)=\phi_0+(t-t_0)/P_0$. For a strictly constant $\dot{P}\neq0$ the same integral admits the closed form $\phi(t)=\phi_0+\dot{P}^{-1}[\ln P(t)-\ln P(t_0)]$, which tends to the linear expression as $\dot{P}\to0$; we verified that the two expressions give identical phase assignments over the period range of interest. The reference time was taken as the first burst arrival on MJD\,59310, $t_0=\mathrm{MJD}\,59310.34008104$; a constant offset in $t_0$ is absorbed by $\phi_0$ and does not affect the search.

Using the phase-folding method, we performed a two-dimensional periodicity search combining the bursts of MJD\,59310 and MJD\,59347. The search range for the period ($P_0$) was set to $1.0$--$10.0$\,s with a step size of $10^{-5}$\,s, and the search range for the period derivative ($\dot{P}$) was set to $1$--$100\times10^{-11}$\,s\,s$^{-1}$ with a step size of $10^{-11}$\,s\,s$^{-1}$. The number of bins used for the phase folding was set to 20. The same search was also applied to all bursts of the entire campaign for comparison.

To estimate the significance of the combined phase-folded signal, we performed Monte Carlo simulations. We generated $10^5$ simulated sets of arrival times following a Poisson distribution, with the number of bursts and the total time span matched to those of the real observation, and applied the same two-dimensional search to each realisation. The periodicity was quantified using the Pearson $\chi^2$ statistic, defined as
\begin{equation}
\label{eqchi2}
\chi^2 = \sum_{j=1}^{n} \frac{(O_j - E_j)^2}{(n-1)E_j},
\end{equation}
where $O_j$ is the observed number of bursts in the $j$-th phase bin, $E_j$ is the expected number under a uniform distribution, and the factor $(n-1)$ in the denominator gives the reduced chi-square.

\begin{figure*}[http!]
\centering
\includegraphics[width=0.45\textwidth]{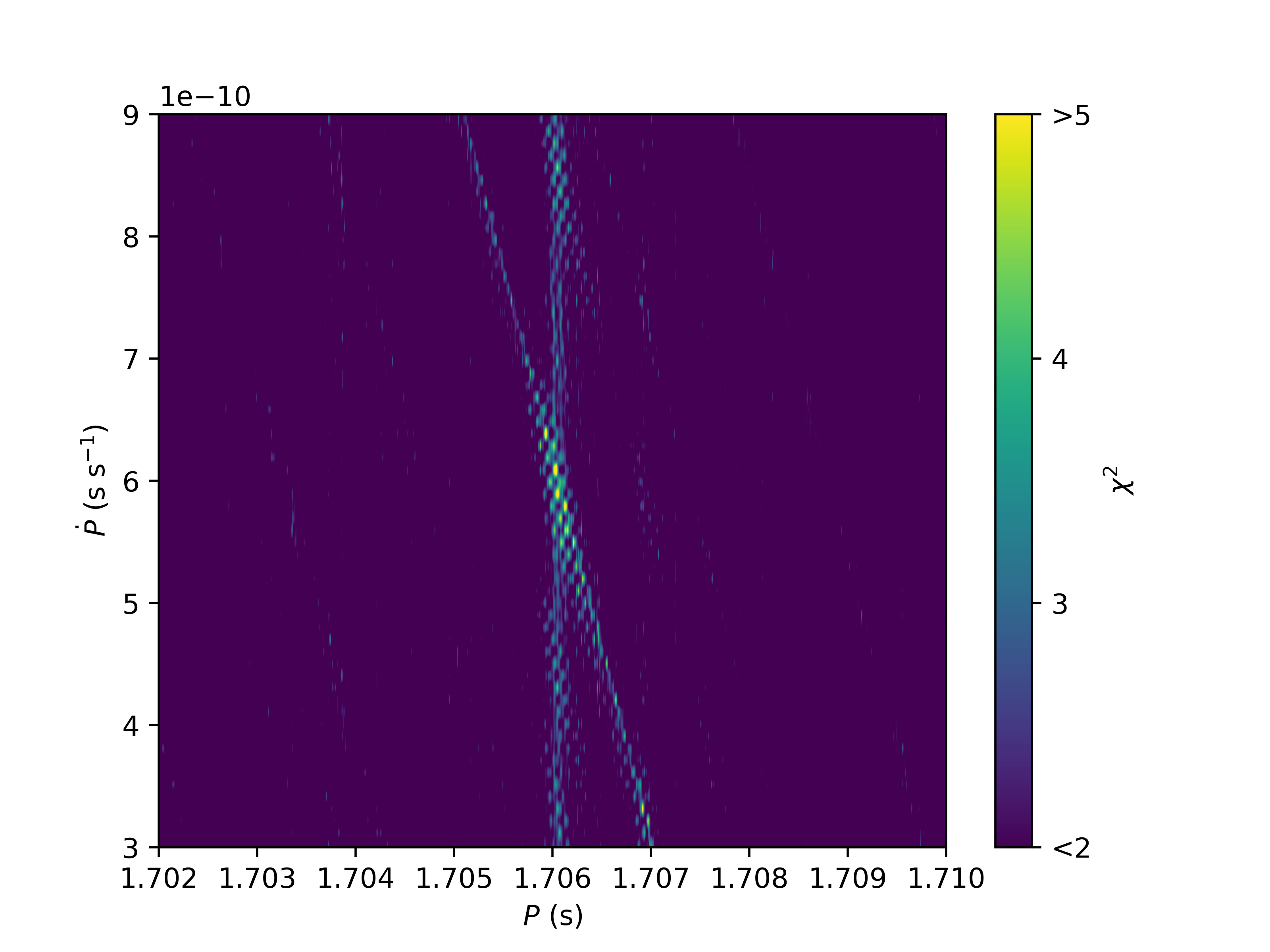}\includegraphics[width=0.5\textwidth, angle=0]{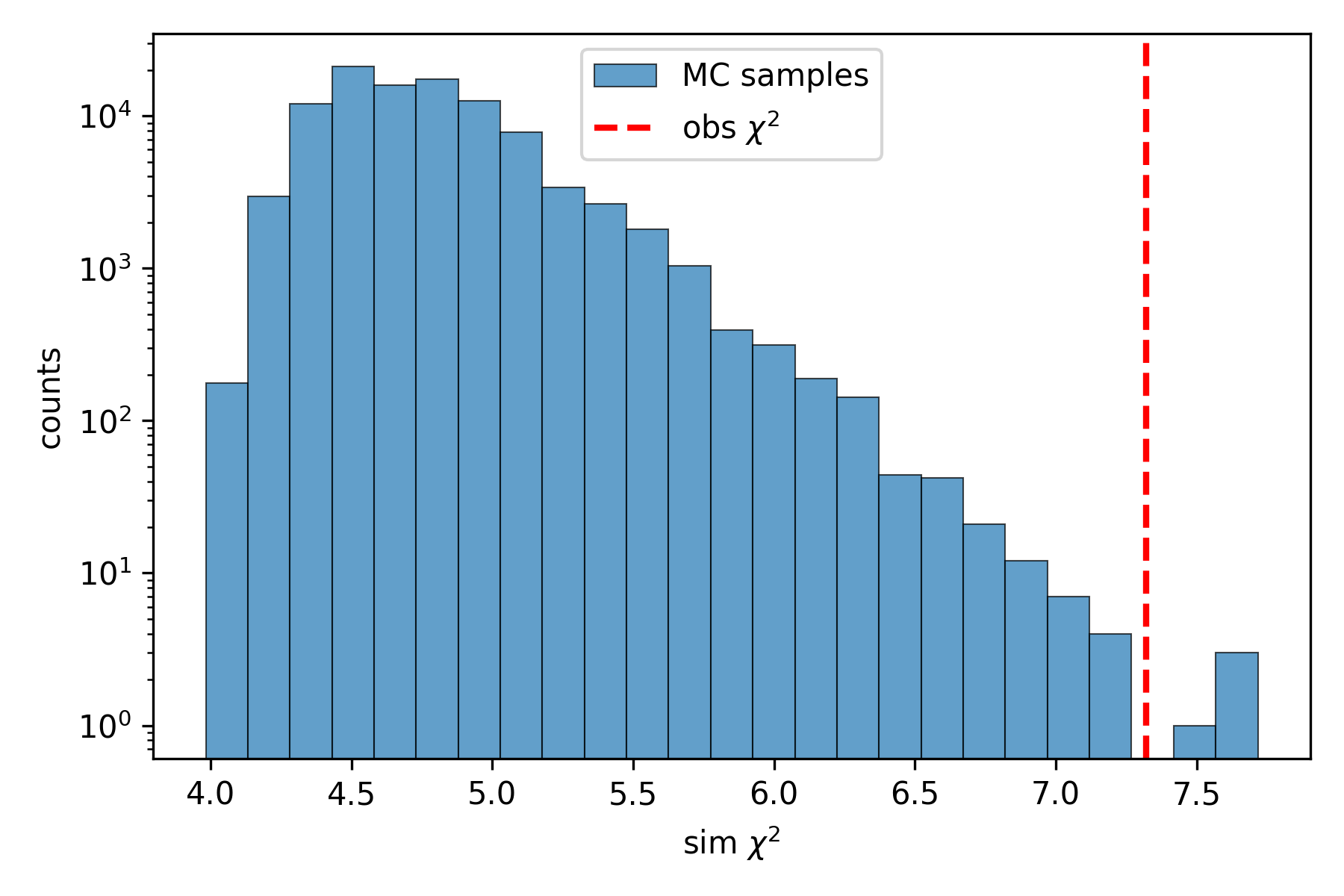}\\
\caption{Left panel: Heatmap of the 2D $P$--$\dot{P}$ periodicity search. Right panel: Distribution of maximum $\chi^2$ values obtained from Monte Carlo samples, with the vertical dotted line indicating the observed $\chi^2$ value.} 
\label{chi2fig}
\end{figure*}

\subsection{Results}

The optimal periodicity solution at the reference epoch $t_0 = \mathrm{MJD}\,59310.34008104$ is $P_0 = 1.70603$\,s and $\dot{P} = 6.1 \times 10^{-10}$\,s\,s$^{-1}$. We then applied the same 2D $P$--$\dot{P}$ periodicity search to each simulated sample. For each simulation, the maximum $\chi^2$ value obtained in the search was recorded. By comparing the observed $\chi^2$ value with the distribution of simulated $\chi^2$ values (left and right panels of Fig.~\ref{chi2fig}), we found that only 4 of the $10^5$ simulated samples produced a value exceeding the observed value, implying a Gaussian significance of $\sim3.9\sigma$.

\begin{figure}[http]
\centering
\includegraphics[width=\columnwidth]{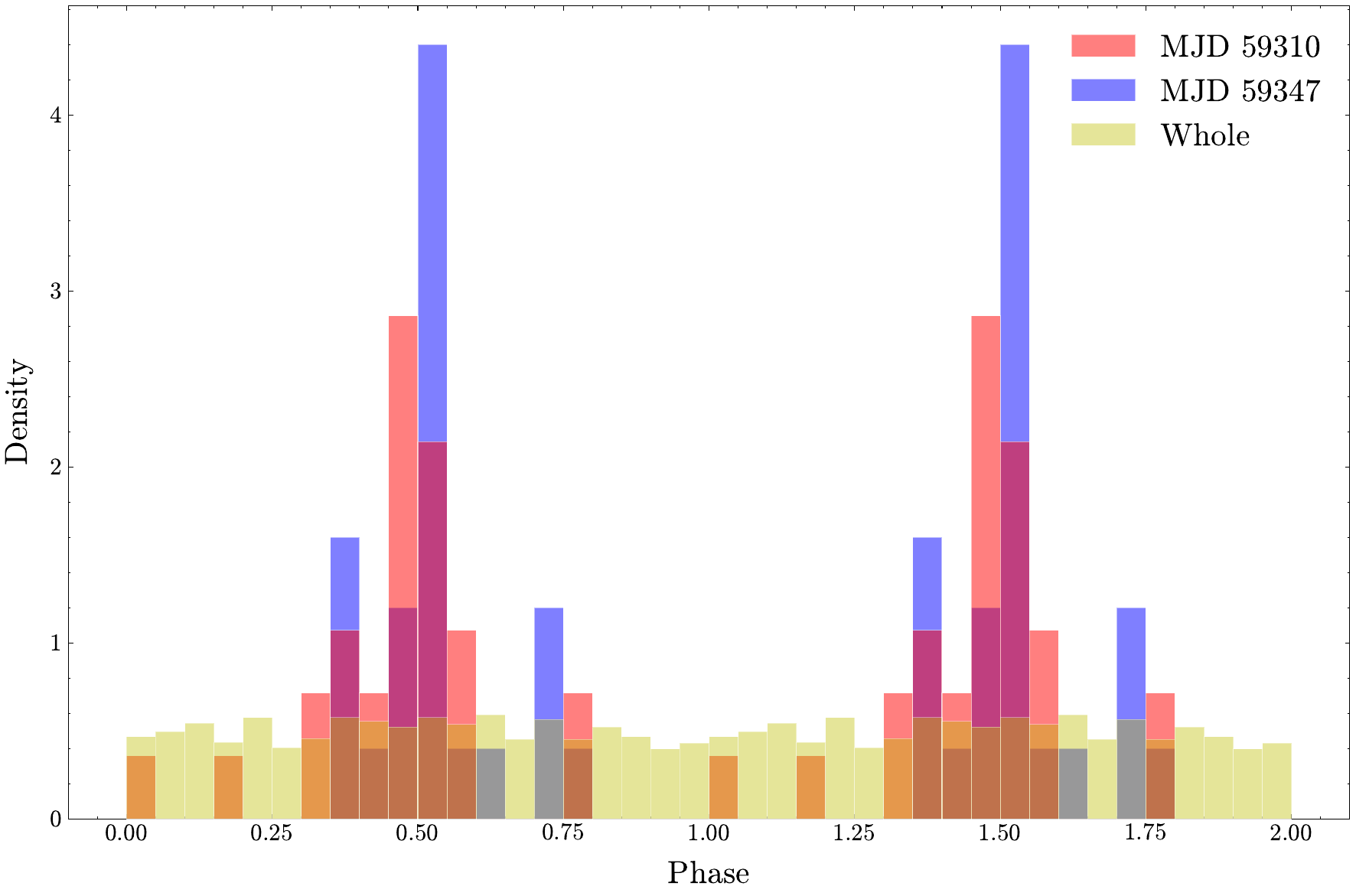}\\
\caption{Comparison of the phase distributions: all bursts versus bursts on MJD 59310 and MJD 59347.}
\label{phases}
\end{figure}

As shown in Fig.~\ref{phases}, the bursts from these two epochs cluster in phase. Through a detailed day-by-day search, \cite{2025arXiv250312013D} identified significant periodic features in only two days, separated by 37 days. Our subsequent coherent period search across both epochs corroborates this result and returns period and period-derivative values consistent with theirs.

\begin{figure}[http!]
\centering
\includegraphics[width=\columnwidth]{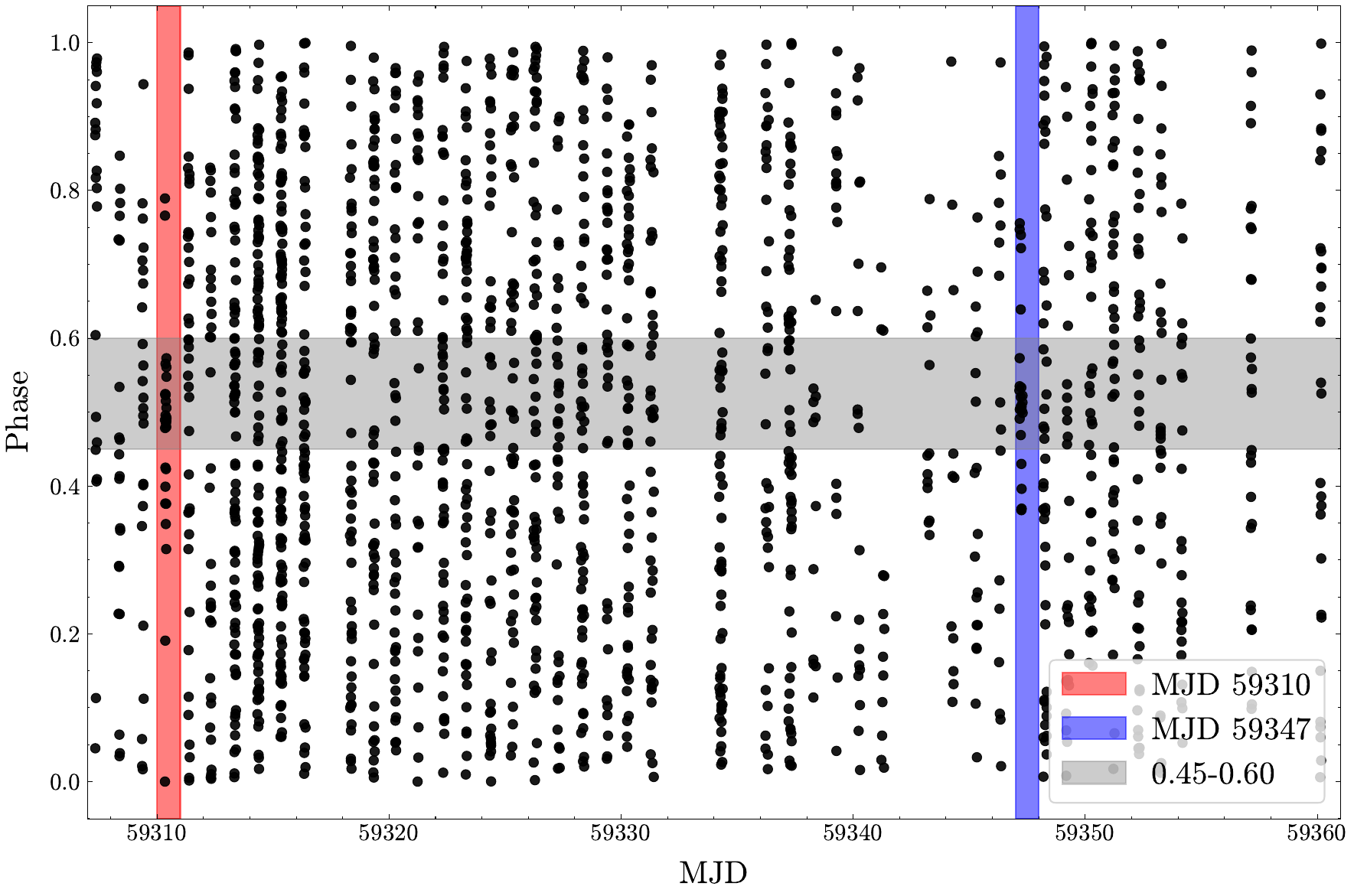}\\
\includegraphics[width=0.5\columnwidth]{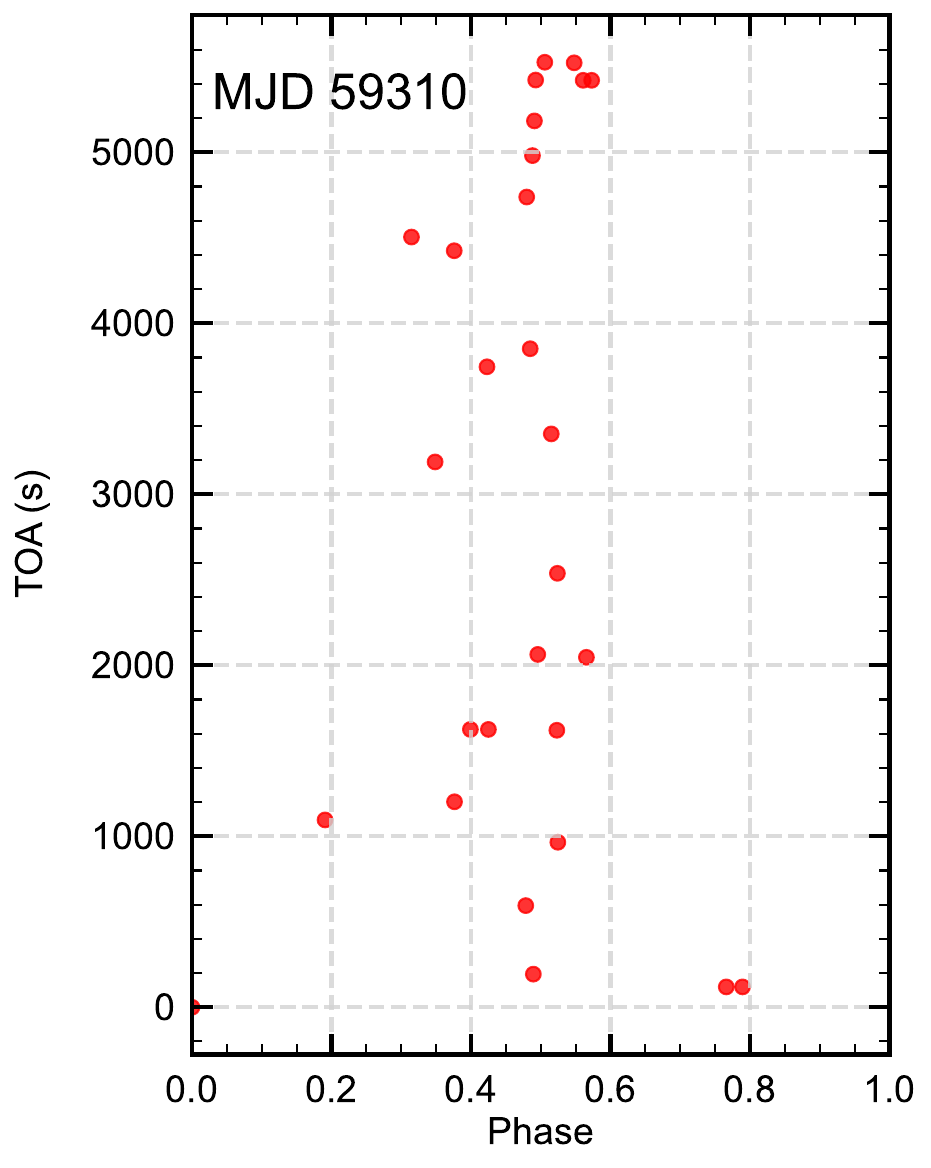}\includegraphics[width=0.5\columnwidth]{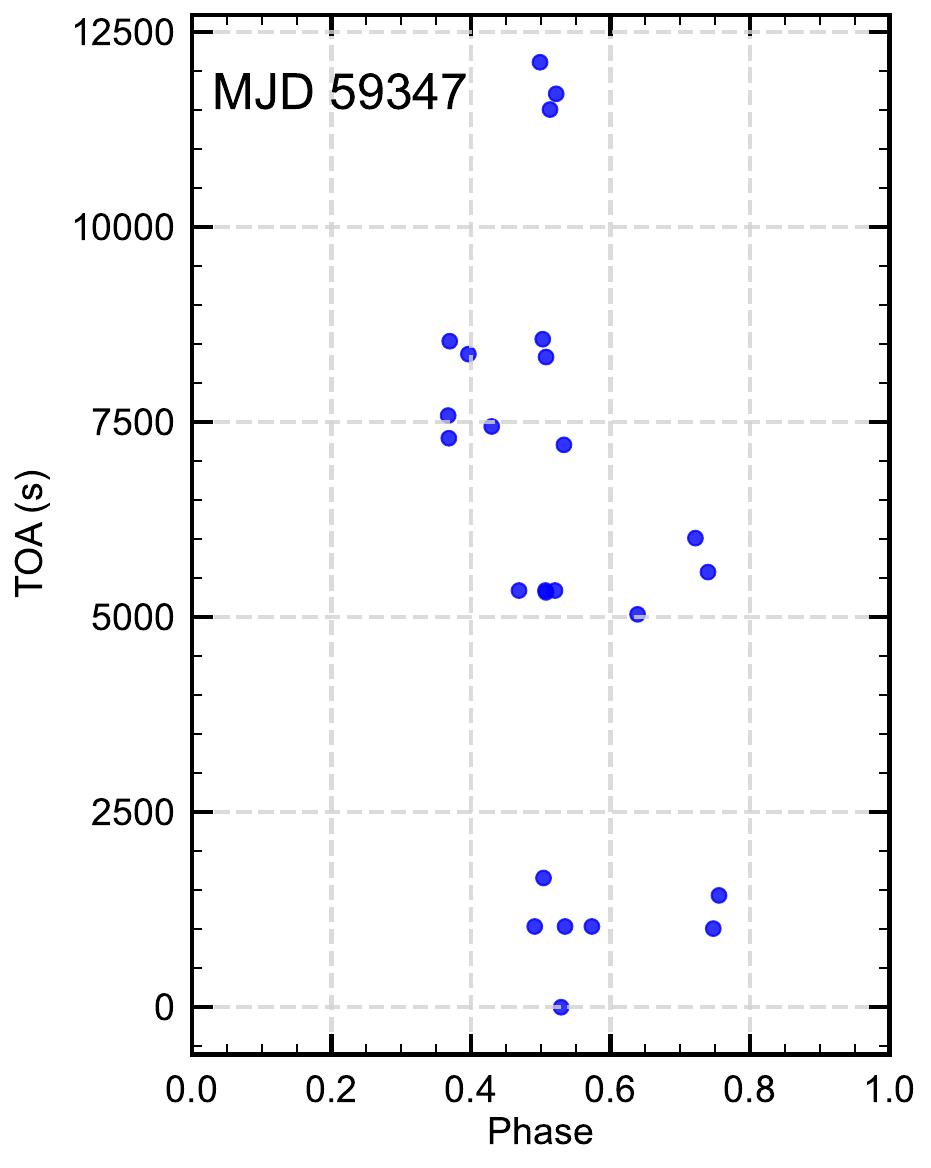}\\
\caption{Bursts from FRB 20201124A detected by FAST plotted on the phase-MJD/TOA plane.}
\label{phase-MJD}
\end{figure}

Fig.~\ref{phase-MJD} reveals the temporal clustering of the bursts detected between MJD\,59310 and MJD\,59347. Within the magnetar paradigm, each cluster likely corresponds to a spatially distinct emission zone. We note that our burst phase calculation (Eq.~\ref{eq2}) does not incorporate the term accounting for the period modulation induced by binary orbital motion. This omission does not compromise the inherent clustering behaviour of the bursts, but rather introduces a systematic phase shift to each burst cluster. The gray-shaded region in Fig.~\ref{phase-MJD} delineates the phase interval corresponding to the statistically significant burst clusters observed on MJD\,59310 and 59347 in Fig.~\ref{phases}. Such a phase-locked emission pattern strongly favors a polar-cap origin, as discussed in Section~\ref{sec:results}.

Finally, we note that the periodicity is recovered on these two days and not on the remaining $\sim50$ days of observation. The physical interpretation of this remarkable fact is discussed in Sections~\ref{sec:results} and~\ref{sec:conclusions}; it is the central question that motivates the magnetar/disk-crossing scenario proposed in this work.

\end{appendix}

\end{document}